\documentclass[prb,aps,twocolumn,reprint,preprintnumbers,amsmath,amssymb]{revtex4-1}
\usepackage{graphicx}
\usepackage{dcolumn}
\usepackage{bm}
\usepackage{amssymb}
\usepackage{epsfig}
\usepackage{color}

\begin{document}
\title{Hybrid superconducting quantum magnetometer}
\author{F. Giazotto}
\email{f.giazotto@sns.it}
\affiliation{NEST Istituto Nanoscienze-CNR  and Scuola Normale Superiore, I-56127 Pisa, Italy}
\author{F. Taddei}
\email{f.taddei@sns.it}
\affiliation{NEST Istituto Nanoscienze-CNR  and Scuola Normale Superiore, I-56127 Pisa, Italy}

\begin{abstract}
A superconducting quantum magnetometer based on magnetic flux-driven modulation of the density of states of a proximized metallic nanowire is theoretically analyzed. With optimized geometrical and material parameters transfer functions up to a few mV$/\Phi_0$ and intrinsic flux noise $\sim10^{-9}\Phi_0/\sqrt{\text{Hz}}$ below 1 K are achievable.
The opportunity to access single-spin detection joined with limited dissipation (of the order of $\sim 10^{-14}$ W) make this magnetometer interesting for the investigation of the switching dynamics of molecules or individual magnetic nanoparticles.
\end{abstract}
\maketitle

\section{Introduction}
The superconducting quantum interference device \cite{Clarke} (SQUID) is recognized as the most sensitive magnetic-flux detector ever realized, and combines   the physical phenomena of Josephson effect \cite{Josephson} and flux quantization \cite{Tinkham} to operate. SQUIDs are nowadays exploited in a variety of physical measurements \cite{gallop,greenberg} with applications spanning, for instance, from pure science to medicine and biology \cite{Clarke,kleiner}. 
Recently, the interest in the development of nanoscale SQUIDs \cite{finkler,hao,troeman} has been motivated by the opportunity to exploit these sensors for the investigation of the magnetic properties of isolated dipoles \cite{ketchen,tilbrook2,tilbrook,huber,cleuziou}, with the ultimate goal to detect one single atomic spin, i.e., one Bohr magneton.  

Here we theoretically analyze  a hybrid superconducting  interferometer 
which exploits the phase dependence of the density of states (DOS) of a proximized metallic nanowire to achieve high sensitivity to magnetic flux.
The operation of a prototype structure based on this principle, the  SQUIPT \cite{squipt}, has been recently reported.
We show that with a careful design transfer functions as large as a few mV$/\Phi_0$ and intrinsic flux noise $\sim 10^{-9}\Phi_0/\sqrt{\text{Hz}}$ can be achieved below 1 K. Limited dissipation joined with the  opportunity to access single-spin detection
make
this structure attractive for the investigation of the switching dynamics of individual magnetic nanoparticles.

The paper is organized as follows. The model of the hybrid superconducting magnetometer is presented in Sec. II. The Josephson and quasiparticle current are calculated in Secs. III and IV, respectively. The flux resolution and device performance are finally presented in Sec. V, where we address briefly the feasibility of this structure as a single-spin detector. Sec. VI is devoted to the Conclusions.
\begin{figure}[t!]
\includegraphics[width=\columnwidth]{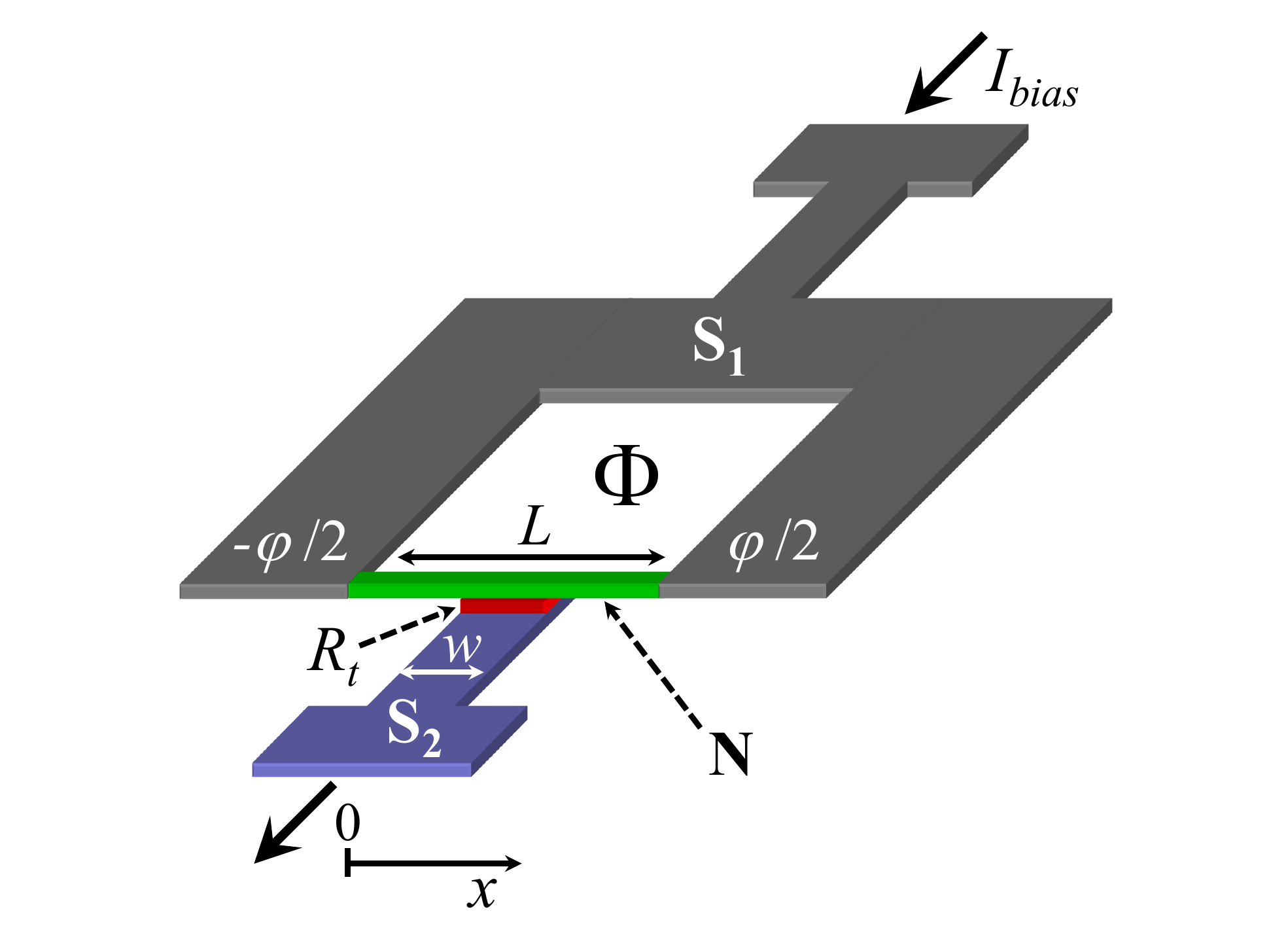}
\caption{\label{fig1} (Color online) (a) Scheme of the device. $L$ is the wire length whereas $w$ is the width of the superconducting tunnel junction (S$_2$) coupled to the middle of the N region. $\varphi$ is the macroscopic quantum phase difference in S$_1$, while $\Phi$ is the magnetic flux threading the loop. Furthermore, $R_t$ is the tunnel junction normal-state resistance and $I_{bias}$ is the current flowing through the structure.
}
\end{figure}
\section{Model}
The interferometer [sketched in Fig. \ref{fig1}(a)] consists of a diffusive normal metal (N) wire of length $L$ in good electric contact (i.e., ideal interface transmissivity) with two superconducting electrodes (S$_1$) which define a ring.  
We assume  the wire transverse dimensions to be much smaller than $L$ so that it can be considered as quasi-one-dimensional.
The contact with S$_1$ induces superconducting  correlations in N through \emph{proximity} effect \cite{proximity,giaz,pot,pet1,pet2} which is responsible for the modification of the wire DOS \cite{gueron}. For lower-transparency NS$_1$ interfaces the proximity effect in the wire will be reduced thus weakening the effects described below.
In addition, a superconducting junction (S$_2$) of width $w$ and normal-state resistance $R_t$ is tunnel-coupled to the middle of the N region. 
The loop geometry allows to change the
phase difference ($\varphi$) across the NS$_1$
boundaries through the application of a magnetic
field which modifies the wire DOS \cite{spivak}
and the transport through the tunnel junction \cite{lesueur,squipt}.

The proximity effect in the wire can be described with the quasiclassical Usadel equations \cite{proximity}. 
The short-junction limit (i.e., for $\Delta_1\ll \hbar D/L^2=E_{Th}$, where $\Delta_1$ is the order parameter in S$_1$, $D$ is the wire diffusion constant, and $E_{Th}$ is the Thouless energy) will be considered in the following, since in such regime the Usadel equations allow an analytic expression for the wire DOS \cite{heikkila} thus simplifying the device analysis. In addition,  the interferometer performance is optimized in this limit as proximity effect in the wire is maximized \cite{spivak,heikkila}.
Assuming a step-function form for the order parameter $\Delta_1$ \cite{selfconsistency}, i.e., constant in S$_1$ and zero in the N wire,
the wire DOS normalized to the DOS at the Fermi level in the absence of proximity effect is given by \cite{heikkila} 
\begin{equation}
N_N(x,\varepsilon,T,\varphi)= \mbox{Re}\left\{\cosh\left[\theta(x,\varepsilon,T,\varphi)\right]\right\},
\end{equation}
where 
\begin{equation}\label{teta}
\theta=\mbox{arcosh}[\alpha(\varepsilon,\varphi,T)\mbox{cosh}[2x\mbox{arcosh}[\beta(\varepsilon,\varphi,T)]]], 
\end{equation}
\begin{equation}
\alpha=\sqrt{\varepsilon^2/[\varepsilon^2-\Delta_1^2(T)\mbox{cos}^2(\varphi/2)]} 
\end{equation}
and
\begin{equation}
\beta=\sqrt{[\varepsilon^2-\Delta_1^2(T)\mbox{cos}^2(\varphi/2)]/[\varepsilon^2-\Delta_1^2(T)]}.
\end{equation}
In the above expressions, $\varepsilon$ is the energy relative to the chemical potential of the superconductors, $T$ is the temperature, 
and $x\in [-L/2,L/2]$ is the spatial coordinate along the wire. 
$N_N$ exhibits a minigap ($\varepsilon_g$) 
\begin{equation}
\varepsilon_g(\varphi)=\Delta_1(T)|\mbox{cos}(\varphi/2)| 
\end{equation}
for $|\varepsilon|\leq \varepsilon_g$ whose amplitude depends on $\varphi$ and is constant along the wire.
In particular, $\varepsilon_g=\Delta_1$ for $\varphi =0$ and decreases by increasing $\varphi$, vanishing at $\varphi=\pi$.
Finally, by neglecting the ring inductance the phase difference becomes $\varphi=2\pi\Phi/\Phi_0$, where $\Phi$ is the total flux through the loop area, and $\Phi_0=2.067\times 10^{-15}$ Wb is the flux quantum. 

\section{Josephson current}

\begin{figure}[t!]
\includegraphics[width=\columnwidth]{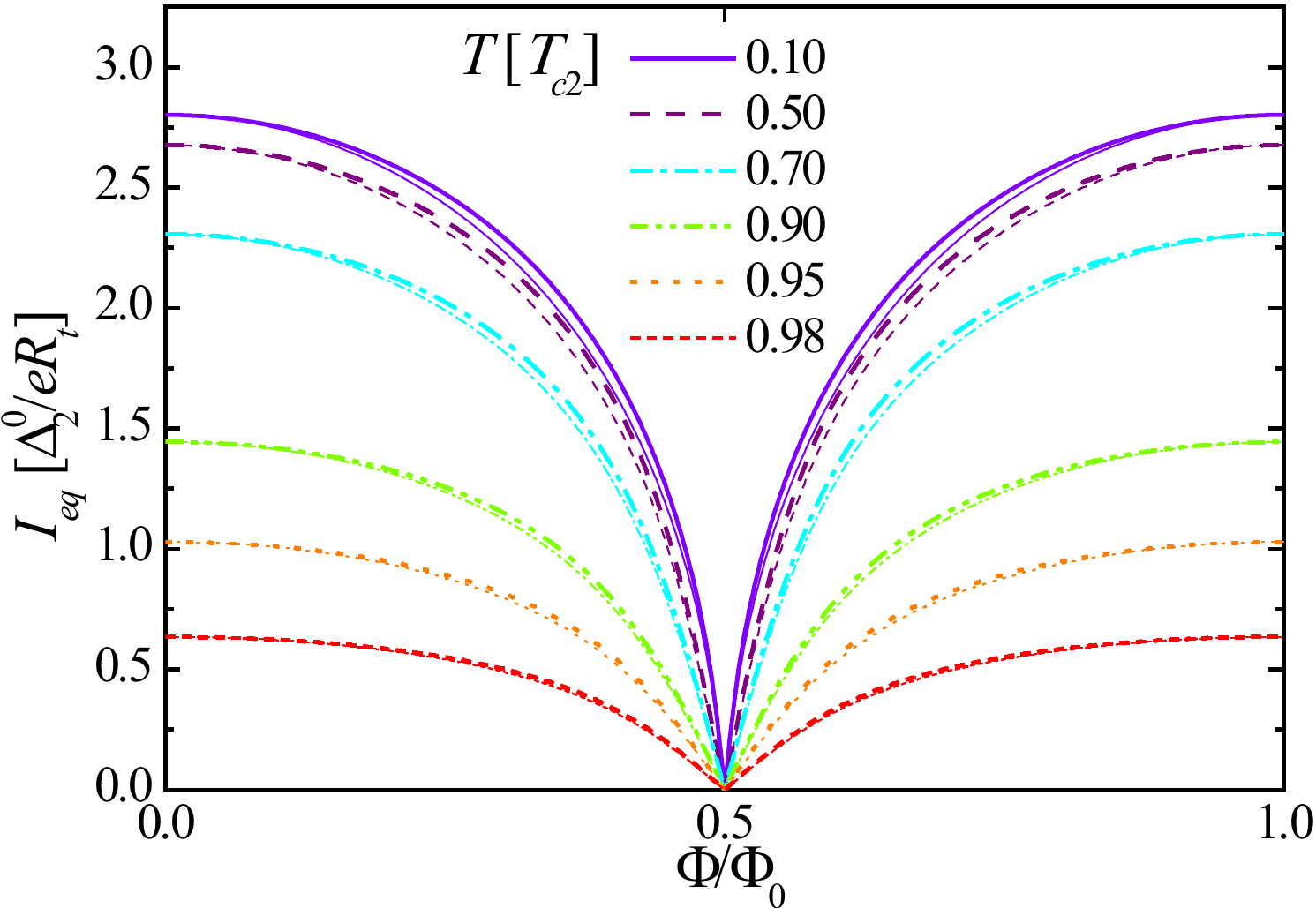}
\caption{\label{fig2} (Color online) Josephson current vs flux ($I_{eq}-\Phi$) characteristics calculated for a few values of temperature $T$ assuming $\Delta_1^0=0.1E_{Th}$ and $\Delta_1^0=4\Delta_2^0$. Thick lines represent the supercurrent calculated using the Ambegaokar-Baratoff formula Eq.~(\ref{Ambe}), whereas the thin lines are the supercurrent calculated for an extended tunnel junction with width $w=L/2$. 
}
\end{figure}
At equilibrium a current through the system $I_{eq}$ can flow thanks to a direct Josephson coupling between the superconducting electrode S$_2$ [with order parameter equal to $\Delta_2(T) e^{i\chi_\text{S}}$ where $\chi_\text{S}$ is the macroscopic phase] and the proximized N wire.
We assume the BCS temperature dependence for $\Delta_{1,2}(T)$ with critical temperature $T_{\text{c1,2}}=\Delta_{1,2}^{0}/(1.764k_{\text{B}})$ where $\Delta_{1,2}^{0}$ is the zero-temperature order parameter in S$_{1,2}$, and
$k_{\text{B}}$ is the Boltzmann constant.
Since the junction NS$_2$ is extended in the $x$-direction one can calculate, in the tunneling limit, $I_{eq}$ using the quasiclassical approach~\cite{nazarov1,nazarov2,proximity} as the following integral 
\begin{eqnarray}
& & I_{eq}(\varphi)=-\frac{1}{8ewR_t}\int_{-w/2}^{w/2} dx \\
& &  \int_{-\infty}^{+\infty} d\epsilon \text{Tr} \{ \sigma_3 [G_{\text{R}}^\text{N}(\epsilon,x,\varphi),G_\text{R}^{\text{S}_2}(\epsilon)]
\tanh (\frac{\epsilon}{2k_BT}) \} ,\nonumber
\end{eqnarray} 
where $\sigma_3$ is the third Pauli matrix, $[\cdot,\cdot]$ represent the commutator, and $e$ is electron charge.
Furthermore, $G_\text{R}^\text{N}(\epsilon,x,\varphi)$ is the position-dependent retarded Green's function on the N wire and $G_\text{R}^{\text{S}_2}(\epsilon)$ is the retarded Green's function of the electrode S$_2$.
They are defined as follows:
\begin{equation}
G_\text{R}^\text{N}(\epsilon,x,\varphi)=\left(
\begin{array}{cc}
\cosh \theta & \sinh \theta e^{i\chi} \\
-\sinh \theta e^{-i\chi} & -\cosh \theta
\end{array}
\right)
\end{equation} 
and
\begin{equation}
G_\text{R}^{\text{S}_2}(\epsilon)=\frac{1}{\sqrt{\epsilon^2-\Delta_2^2}}
\left(
\begin{array}{cc}
\epsilon & \Delta_2 e^{i\chi_\text{S}} \\
-\Delta_2 e^{-i\chi_\text{S}} & -\epsilon
\end{array}
\right) ,
\end{equation}
where $\theta$ is defined in Eq.~(\ref{teta}) and
\begin{equation}\label{chi}
\chi=-\mbox{arctan}[\gamma(\varepsilon,\varphi,T)\mbox{tanh}[2x\mbox{arcosh}[\beta(\varepsilon,\varphi,T)]]], 
\end{equation}
with
\begin{equation}
\gamma=\sqrt{[\varepsilon^2-\Delta_1^2(T)\mbox{cos}^2(\varphi/2)]}/[\Delta_1(T)\mbox{cos}(\varphi/2)].
\end{equation}
Note that the supercurrent $I_{eq}$ depends on the phase $\chi_S$ of the order parameter in the electrode S$_2$. We are interested in the critical current that we determine by fixing $\chi_S$ such that it gives the maximum supercurrent for a given value of $\varphi$.
In Fig.~\ref{fig2} the equilibrium critical current is plotted, for different values of temperature, as a function of $\varphi$ in units of $\Delta_2^0/(eR_t)$, assuming $\Delta^0_1=0.1E_{Th}$, $\Delta_1^0=4\Delta^0_2$ and $w=L/2$.
For the sake of comparison we also plot the supercurrent calculated through the Ambegaokar-Baratoff~\cite{ab} formula, relative to a point-like NS$_2$ junction:
\begin{eqnarray}\label{Ambe}
& & I_{eq}^\text{AB}=\frac{\pi\varepsilon_g(\varphi)\Delta_2(T) k_{\text{B}}T}{eR_t}\times \\
& & \sum_{l=0,\pm1,\cdots}\frac{1}{\sqrt{[\omega_l^2+\varepsilon_g^2(\varphi)][\omega_l^2+\Delta_2^2(T)]}} ,\nonumber
\end{eqnarray}
where $\omega_l=\pi k_{\text{B}}T(2l+1)$.
Interestingly, for our choice of parameters the difference between $I_{eq}$ and $I_{eq}^\text{AB}$ is hardly noticeable: the lateral spatial extension of the NS$_2$ junction along $x$ plays a marginal role.
As a matter of fact the supercurrent turns out to be negligible in the experiments reported in Refs.~\onlinecite{squipt,meschke}.
The reason for such Josephson current suppression is presently unclear. One possibility is that the system is brought out of equilibrium by voltage fluctuations which might originates from the measuring circuit. Such voltage fluctuations will drop mostly across the tunneling barrier, being the most resistive component of the system. As a result, the Josephson current will oscillates at high frequency, hindering the possibility of detection.
This fact is fortunate, since a supercurrent might prevent a correct voltage read-out of the device, which is not observed experimentally.

\section{Quasiparticle current}

The current through the tunnel junction biased at voltage $V$ is therefore dominated by the quasiparticles, and can be written as \cite{Tinkham}
\begin{equation}
I=\frac{1}{ewR_t}\int_{-w/2}^{w/2}dx\int d\varepsilon N_N(x,\varepsilon,\varphi)N_{S2}(\tilde{\varepsilon})F(\varepsilon,\tilde{\varepsilon}),
\end{equation} 
where
\begin{equation}
N_{S2}(\varepsilon,T)=\frac{|\varepsilon|}{\sqrt{\varepsilon^2-\Delta_2(T)^2}}\Theta[\varepsilon^2-\Delta_2(T)^2] 
\end{equation}
is the normalized DOS of the S$_2$ electrode,
$\tilde{\varepsilon}=\varepsilon-eV$,
$\Theta(y)$ is the Heaviside step function, $F(\varepsilon,\tilde{\varepsilon})=[f_0(\tilde{\varepsilon})-f_0(\varepsilon)]$,
and $f_0(\varepsilon)=[1+\mbox{exp}(\varepsilon/k_BT)]^{-1}$ is the Fermi-Dirac energy distribution.
In the following we set $\Delta_2^0=200\,\mu$eV and  $\Delta_{1}^0=4\Delta_2^0=800\,\mu$eV as representative values for a structure exploiting aluminum (Al) and vanadium (V) as superconductors \cite{pascual,quaranta}, respectively, $w=L/2$ and $R_t=5$ M$\Omega$.
\begin{figure}[t!]
\includegraphics[width=\columnwidth]{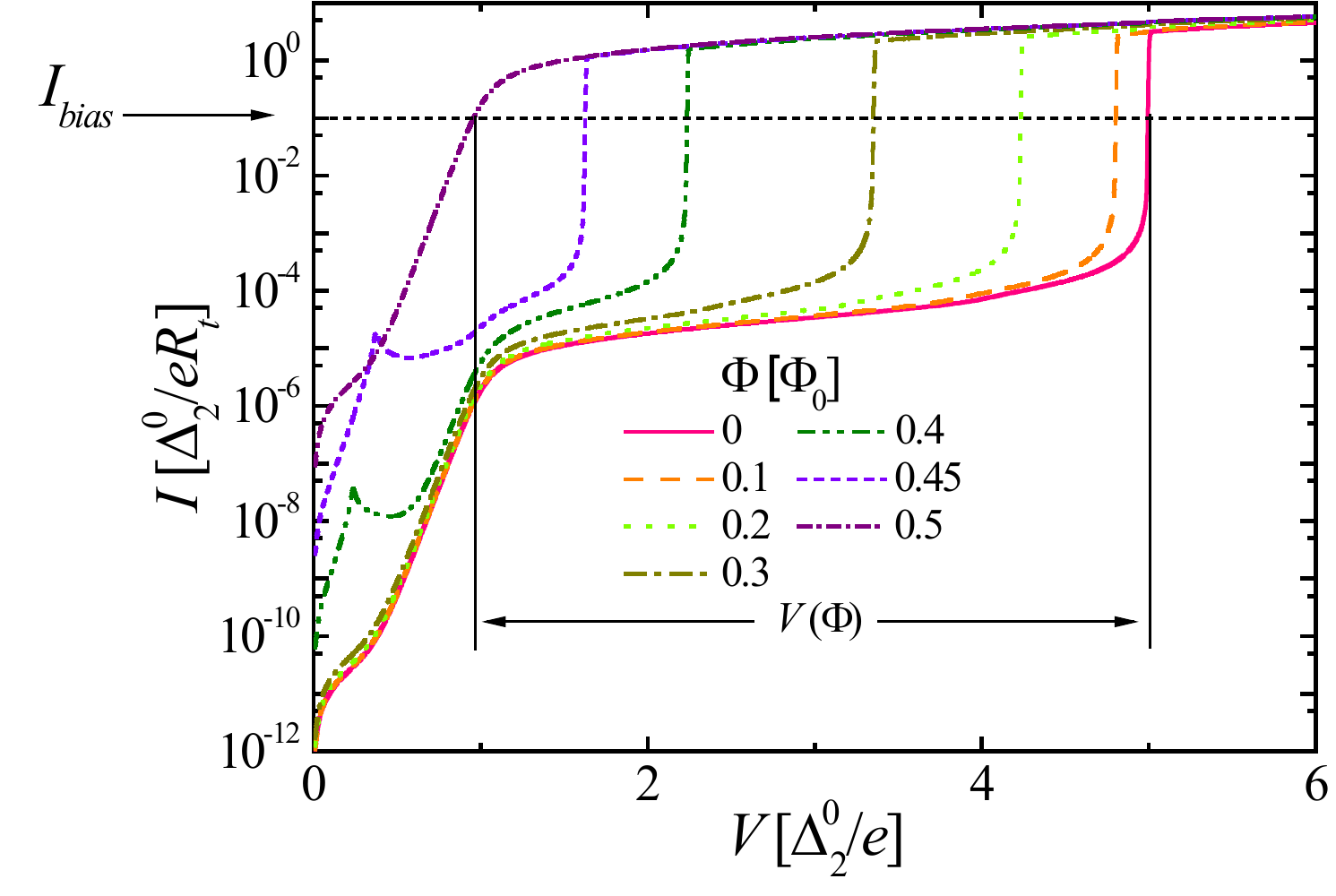}
\caption{\label{fig3} (Color online) Interferometer quasiparticle current vs voltage  ($I-V$) characteristics calculated for a few values of $\Phi$ at $T=0.1 T_{c2}$. $T_{c2}$ is the critical temperature of S$_2$, $I_{bias}$ is the current flowing through the device and $V(\Phi)$ is the resulting voltage modulation.
}
\end{figure}

Figure 3 shows the interferometer current vs voltage ($I$-$V$) characteristics calculated at $T=0.1T_{c2}$ for different values of the applied flux $\Phi$ \cite{gamma}. 
In particular, for $\Phi=0$ the $I-V$ characteristic resembles that typical of a superconductor-insulator-superconductor junction (i.e., S$_1$IS$_2$ where I denotes an insulator) where the minigap in the wire is maximized [i.e., $\varepsilon_g=\Delta_1(T)$], and the onset for large quasiparticle current occurs at \cite{Tinkham} $V=[\Delta_1(T)+\Delta_2(T)]/e$. 
For $\Phi=\Phi_0/2$ the characteristic is similar to that of a normal metal-insulator-superconductor junction (i.e., NIS$_2$) with $\varepsilon_g$ suppressed. The curves show a peak at $V=|\Delta_1(T)-\Delta_2(T)|/e$ which corresponds to the singularity appearing in the tunneling characteristic between different superconductors \cite{Tinkham}. 
In a current-biased setup 
the interferometer operates as a flux-to-voltage transducer providing a voltage response $V(\Phi)$ that depends on the bias current $I_{bias}$ fed through the tunnel junction [see Fig. 3].
For any $I_{bias}$, $V(\Phi)$ is determined by solving the equation $I_{bias}-I=0$.  

\begin{figure}[t!]
\includegraphics[width=\columnwidth]{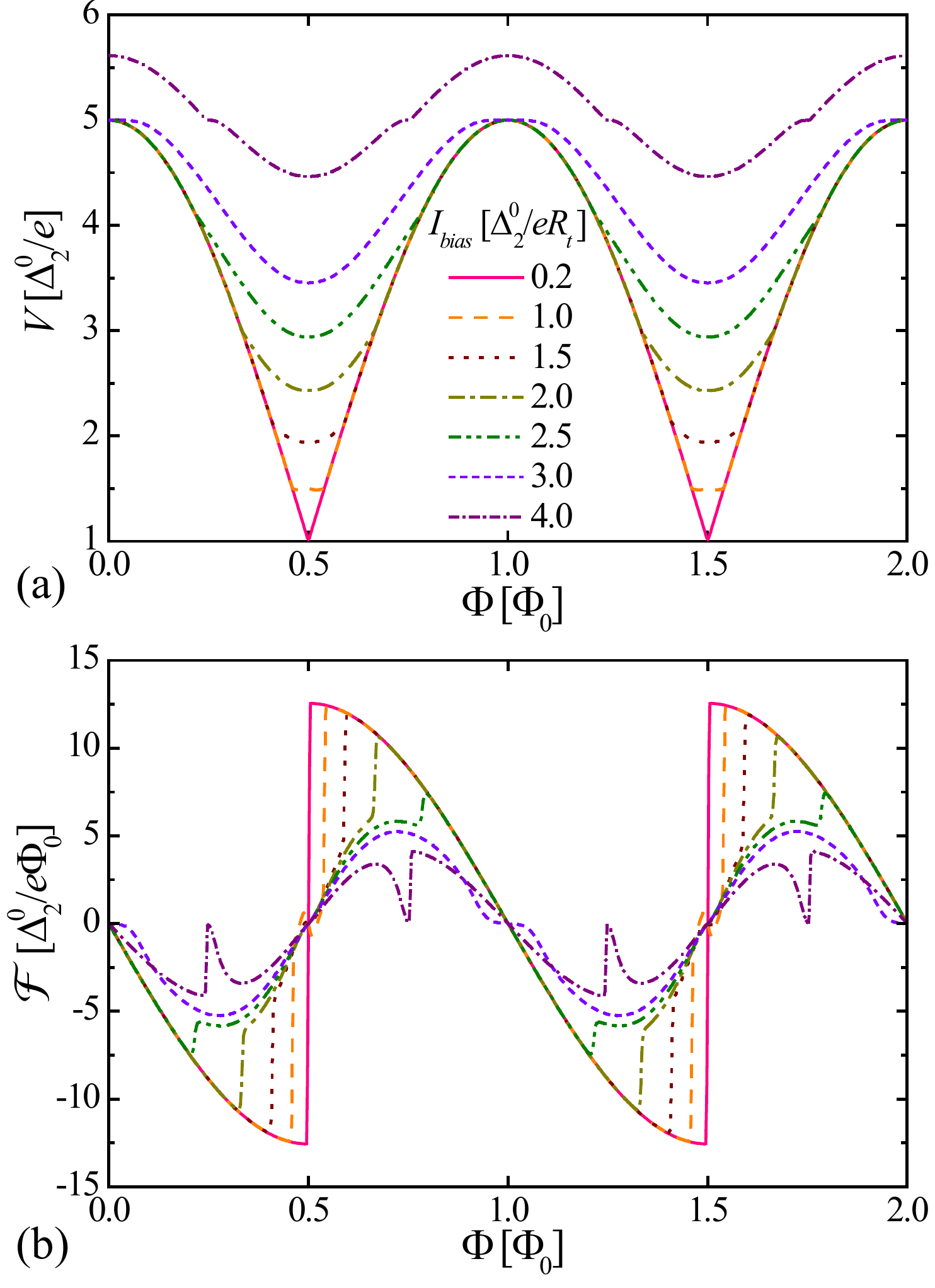}
\caption{\label{fig4} (Color online) (a) $V$ vs $\Phi$ calculated for several bias currents $I_{bias}$ at $T=0.1T_{c2}$. (b) $\mathcal{F}$ vs $\Phi$ calculated for the same $I_{bias}$ values and $T$ as in (a).
}
\end{figure}
Figure 4(a) shows $V(\Phi)$ at $T=0.1T_{c2}$ calculated for several $I_{bias}$ values. $V(\Phi)$ turns out to be maximized at the lower bias currents where the voltage swing obtains values as large as $4\Delta_2^0/e$, whereas it is gradually reduced by increasing $I_{bias}$.  The interferometer performance is thus improved at low $I_{bias}$.  

An important figure of merit of the interferometer is represented by the flux-to-voltage transfer function \cite{Clarke} 
\begin{equation}
\mathcal{F}(\Phi)=\frac{\partial V}{\partial \Phi}
\end{equation}
which is shown in Fig. 4(b) for the same $I_{bias}$ values as in panel (a). In particular, $\mathcal{F}$ as large as $\simeq 12.5\Delta_2^0(e\Phi_0)^{-1}$ can be obtained at the lowest currents, whereas it is gradually suppressed at higher $I_{bias}$. 
\begin{figure}[t!]
\includegraphics[width=\columnwidth]{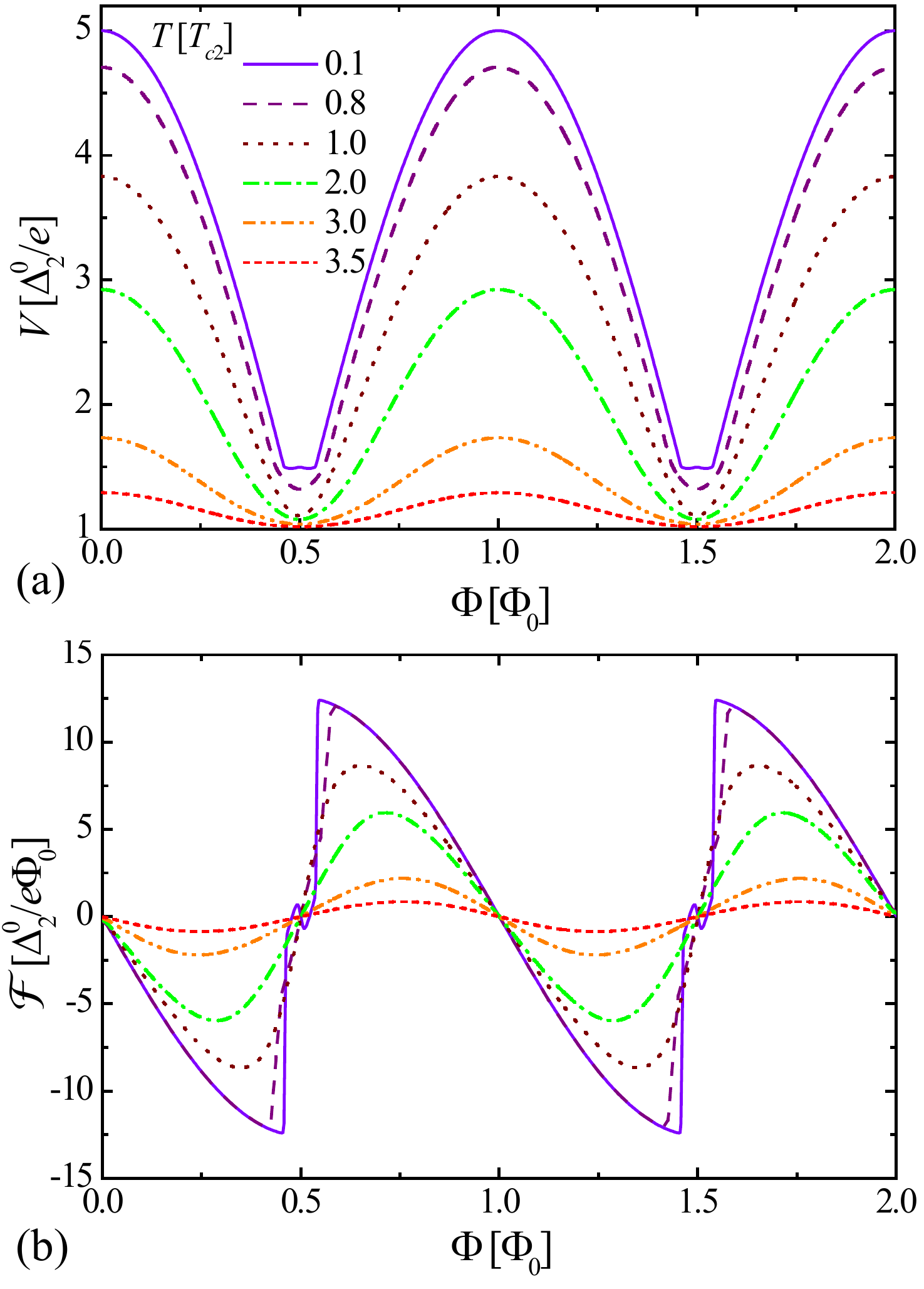}
\caption{\label{fig5} (Color online) (a) $V$ vs $\Phi$ calculated for a few temperatures at $I_{bias}=1.0\Delta_2^0/eR_t$. (b) $\mathcal{F}$ vs $\Phi$ calculated for the same $T$ values and $I_{bias}$ as in (a).
}
\end{figure}

The role of the temperature is shown in Fig. 5(a) which displays $V(\Phi)$ calculated for several $T$ values at $I=1.0\Delta_2^0/(eR_t)$.
An increase in $T$ leads to a reduction of $V(\Phi)$ as well as to a suppression and smearing of the voltage swing. 
This directly reflects on the transfer function, as displayed in Fig. 5(b). We note that even at $T=T_{c,2}$, i.e., when S$_2$ is driven into the normal state,
$\mathcal{F}$ as large as $\simeq 8.6\Delta_2^0(e\Phi_0)^{-1}$ can be achieved.
It follows that voltage swings up to 0.8 mV and $\mathcal{F}$ as large as 2.5 mV/$\Phi_0$ can be achieved with the suggested materials combination for $T\lesssim1$ K.

\section{Noise and device performance}

We now turn on discussing the noise properties of the interferometer. 
In the actual current-biased setup an important quantity is represented by the voltage noise spectral density ($S_V$) defined as 
\begin{equation}
S_V=R_d^2S_I, 
\end{equation}
where $R_d=\partial V/\partial I$ is the tunnel junction dynamic resistance, and  $S_I$ is the current noise spectral density (shot noise) given by \cite{golubev}
\begin{equation}
S_I=\frac{2}{wR_t}\int_{-w/2}^{w/2}dx\int d\varepsilon N_N(x,\varepsilon,\Phi)N_{S2}(\tilde{\varepsilon})M(\varepsilon,\tilde{\varepsilon}),
\end{equation}
where 
\begin{equation}
M(\varepsilon,\tilde{\varepsilon})=f_0(\tilde{\varepsilon})[1-f_0(\varepsilon)]+f_0(\varepsilon)[1-f_0(\tilde{\varepsilon})]. 
\end{equation}
The  intrinsic flux noise per unit bandwidth of the interferometer ($\Phi_{ns}$) is related to the voltage noise spectral density as \cite{Clarke} 
\begin{equation}
\Phi_{ns}=\frac{\sqrt{S_V}}{|\mathcal{F}(\Phi)|}. 
\end{equation}
Note that $\Phi_{ns}\propto \sqrt{R_t}$, as $S_V\propto R_t$ and $\mathcal {F}(\Phi)$ is independent of tunnel junction resistance.

\begin{figure}[t!]
\includegraphics[width=\columnwidth]{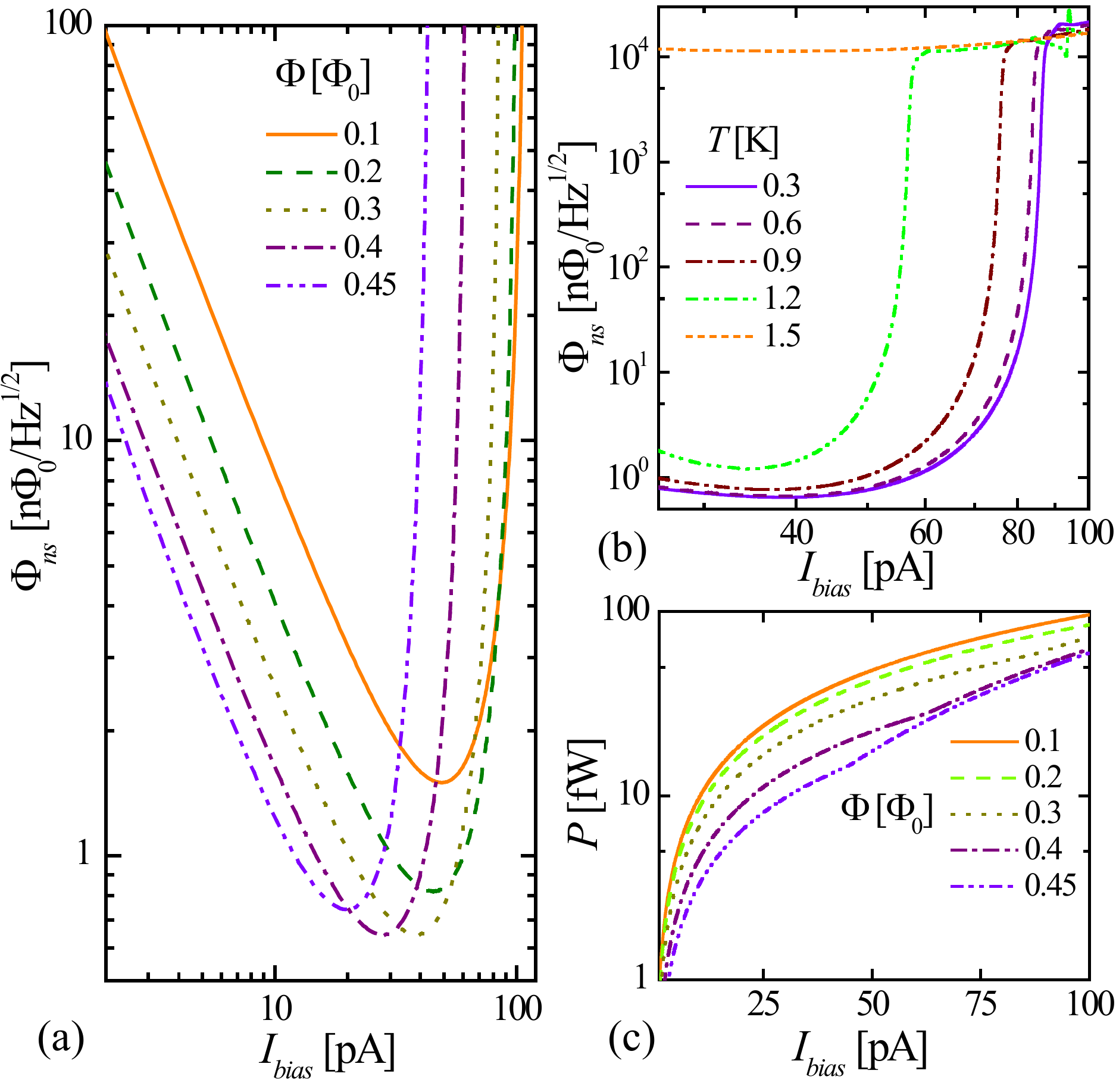}
\caption{ (Color online) (a) $\Phi_{ns}$ vs $I_{bias}$ calculated for different $\Phi$ values at $T=0.3$ K.
(b) $\Phi_{ns}$ vs $I_{bias}$ for $\Phi=0.3\Phi_0$ calculated at different temperatures.
(c)  $P$ vs $I_{bias}$ calculated for different  $\Phi$ values at $T=0.3$ K. In all calculations we set $\Delta_2^0=200\,\mu$eV, $\Delta_1^0=800\,\mu$eV, and $R_t=5$ M$\Omega$.
}
\label{fig6}
\end{figure}

Figure \ref{fig6}(a) shows $\Phi_{ns}$ versus $I_{bias}$ for several flux values at $T=300$ mK. 
$\Phi_{ns}$ is a non-monotonic function of $I_{bias}$ with a minimum which depends, for each $\Phi$, on the bias current. 
In particular,  an increase in $\Phi$ leads to a general reduction of $\Phi_{ns}$ at low $I_{bias}$, while
its minimum moves toward lower bias current.  
We stress that $\Phi_{ns}$ as low as $10^{-9}\,\Phi_0/\sqrt{\text{Hz}}$ or better can be achieved at this temperature in the $\sim 10...80$ pA range for suitable values of $\Phi$. 
This good flux sensitivity stems from the low shot noise $S_I$ (which is peculiar to  all-superconducting tunnel junctions) together with a small $R_d$ at the biasing point and large $\mathcal{F}(\Phi)$.

The temperature dependence is displayed in Fig. \ref{fig6}(b) where $\Phi_{ns}$ vs $I_{bias}$ is plotted for different $T$ values at $\Phi=0.3\Phi_0$.
Notably, the minimum of $\Phi_{ns}$ turns out to be quite insensitive to the temperature up to $\simeq 900$ mK.  
Then, higher $T$ yields to a reduction of the current window suitable for high flux sensitivity and to an overall enhancement of $\Phi_{ns}$. 
Furthermore, for $T\geq T_{c2}$ [see the line corresponding to $T=1.5$ K in  Fig. 6(b)] $\Phi_{ns}$ is significantly degraded in the whole $I_{bias}$ range. This emphasizes the effectiveness of a superconducting tunnel probe for a drastic suppression of $\Phi_{ns}$.  

The impact of dissipation $P=VI$ is displayed in Fig.~\ref{fig6}(c) which shows $P$ vs $I_{bias}$ for different $\Phi$ values at $T=0.3$ K. 
$P$ can largely change by varying $\Phi$ and $I_{bias}$ as well.
In particular, in the $\sim 10...80$ pA current range, $P$ can vary from a few fW to some tens of fW. 
Such a small power has the additional advantage to prevent  substantial electron heating in the N wire \cite{rmp}.  By contrast, in conventional SQUIDs dissipation is typically  from two to five orders of magnitude larger \cite{Clarke,kleiner}. As $P\propto R_t^{-1}$, dissipation can be tailored by choosing a proper value of the tunnel junction resistance.

For a correct operation of the interferometer the two following conditions should be fulfilled: i) $2\pi I_c^0\mathcal{L}_G/\Phi_0\lesssim 1$ \cite{Tinkham} (where $I_c^0$ is the zero-temperature critical current of the S$_1$NS$_1$ Josephson junction, and $\mathcal{L}_G$ is the loop geometric inductance), and  ii) $\mathcal{L}_k^{S_1}\ll \mathcal{L}_k^{N}$ \cite{lesueur,squipt} [where $\mathcal{L}_k^{S_1,N}\simeq \hbar R_{S_1,N}/\pi \Delta_1^0$ is the kinetic inductance \cite{Tinkham}, and $R_{S_1,N}$ is the normal-state resistance of S$_1$(N)]. 
Condition i), where $I_c^0=0.66\pi\Delta_1^0/eR_N$ \cite{heikkila}, ensures to avoid magnetic hysteresis whereas ii), which is equivalent to $R_{S_1}\ll R_N$, ensures that the phase difference set by $\Phi$ drops entirely at the wire ends thus allowing a full modulation of its DOS. 
As an additional set of parameters we choose a silver (Ag) wire with $L=80$ nm, cross section $\mathcal{A}=30\times10$ nm$^2$,  and $D=0.02$ m$^2$s$^{-1}$ which yield $R_N=L/(\mathcal{A}\nu_F e^2D)\simeq 5.2\,\Omega$, where $\nu_F=1\times 10^{47}$ J$^{-1}$m$^{-3}$ is the DOS at the Fermi level in Ag, $\Delta_1/E_{Th}\simeq 0.3$, and $I_c^0\simeq 318$ $\mu$A. 
By choosing, for instance, a circular washer geometry \cite{Clarke} with $2r=150$ nm as internal diameter and external radius $\mathcal{R}\gg r$ we get $\mathcal{L}_G=2\mu_0 r\approx 0.19$ pH, where $\mu_0$ is the vacuum permeability, so that $2\pi I_c^0\mathcal{L}_G/\Phi_0\approx 0.18$. 
Condition ii) can be fulfilled as well by choosing a suitable washer thickness and $\mathcal{R}$. 

As $\mathcal{L}_G$ has to be kept small to satisfy condition i) it follows that the present structure could be suitable for the measurement of the magnetic properties of small isolated samples. 
In this context, the magnetometer sensitivity ($\mathcal{S}_n$) to an isolated magnetic dipole placed at the center of the loop is approximtely given by \cite{ketchen,tilbrook} 
\begin{equation}
\mathcal{S}_n=\frac{2r\Phi_{ns}}{\mu_0\mu_B}, 
\end{equation}
where $\mu_B$ is the Bohr magneton. 
With our choice for $r$ and by coupling the device to a cryogenic voltage preamplifier \cite{kiviranta} (which we assume dominates the voltage noise) with $\sqrt{S_V^{pre}}\simeq 0.1$ nV$/\sqrt{\text{Hz}}$ yields a total flux noise $\Phi_{ns}^{tot}=\frac{\sqrt{S_V^{pre}}}{\text{max}|\mathcal{F}(\Phi)|}\simeq 40$ n$\Phi_0/\sqrt{\text{Hz}}$, leading to
$\mathcal{S}_n\approx 1$ atomic spin$/\sqrt{\text{Hz}}$ below 1 K. Furthermore, the best achievable energy resolution \cite{Clarke} would be $\mathcal{E}=\frac{(\Phi_{ns}^{tot})^2}{2\mathcal{L}_G}\simeq 170\hbar$.

\section{Conclusions}
In summary, we have theoretically investigated a hybrid superconducting magnetometer whose operation is based on magnetic flux-driven modulation of the density of states of a proximized metallic nanowire.	In particular, we have shown that with suitable geometrical and material parameters the interferometer can provide large transfer functions (i.e., of the order of a few mV/$\Phi_0$) and intrinsic flux noise down to a few n$\Phi_0/\sqrt{\text{Hz}}$ below 1K. Furthermore, joined with limited power dissipation, the structure has the potential for the realization of sensitive magnetometers for the investigation of the switching dynamics of small spin populations.

\acknowledgments

We acknowledge L. Faoro, R. Fazio, P. Helist$\ddot{\text{o}}$, L. B. Ioffe, M. Kiviranta, M. Meschke, Yu. V. Nazarov, J. P. Pekola, and S. Pugnetti for fruitful discussions. The FP7 program ``MICROKELVIN'' and the EU project ``SOLID'' are acknowledged for partial financial support.


\begin{thebibliography}{99}
\bibitem{selfconsistency}
We note that although a self-consistent calculation could be performed by including suppression of the order parameter in the superconducting ring at the NS$_1$ boundaries due to inverse proximity effect, in the considered limit $\Delta_1 \ll E_{Th}$
this would lead  to an almost negligible variation of the structure response \cite{proximity,lesueur,cuevas}. In particular, we expect this effect to be small in an experimental situation as well by making the cross section of S$_1$ much larger than that of N.
We have therefore chosen not to include self-consistency which might be, however, useful for the fine analysis of a real device.
\bibitem{gamma}
For the calculations we added in $N_{N,S_2}(\varepsilon)$ a small imaginary part to the energy to account for smearing, i.e.,  $\varepsilon \rightarrow \varepsilon +i\Gamma_{N,S_2}$ where $\Gamma_{N,S_2} =10^{-5}\Delta_{1,2}^0$ \cite{env}.
\end{thebibliography}
\end{document}